\documentclass[pra,amsfonts,amsmath,twocolumn,superscriptaddress,nofootinbib]{revtex4}

\usepackage{amssymb,color,paralist,amsmath,epsfig}
\usepackage{graphicx}
\usepackage{comment}
\usepackage{amsfonts}
\usepackage{relsize}
\usepackage{nicefrac,esint}
\usepackage{float}

\usepackage[colorlinks,citecolor=blue,linktoc=all,linkcolor=cyan]{hyperref}

\newcommand{\beq}{\begin{equation}}
\newcommand{\eeq}{\end{equation}}

\newcommand{\ber}{\begin{eqnarray}}
\newcommand{\eer}{\end{eqnarray}}

\begin{document}

\title{Hyperfine splitting in ordinary and muonic hydrogen}

\author{Oleksandr Tomalak}
\affiliation{Institut f\"ur Kernphysik and PRISMA Cluster of Excellence, Johannes Gutenberg Universit\"at, Mainz, Germany}
%\affiliation{Institut f\"ur Kernphysik \& Cluster of Excellence PRISMA, Johannes Gutenberg Universit\"at Mainz, 55128 Mainz, Germany}
%\affiliation{PRISMA Cluster of Excellence, Johannes Gutenberg-Universit\"at,  Mainz, Germany}
%\affiliation{Department of Physics, Taras Shevchenko National University of Kyiv, Ukraine}

%\author{Marc Vanderhaeghen}
%\affiliation{Institut f\"ur Kernphysik \& Cluster of Excellence PRISMA, Johannes Gutenberg Universit\"at Mainz, 55128 Mainz, Germany}
%\affiliation{PRISMA Cluster of Excellence, Johannes Gutenberg-Universit\"at,  Mainz, Germany}

\date{\today}

\begin{abstract}
We provide an accurate evaluation of the two-photon exchange correction to the hyperfine splitting of S energy levels in muonic hydrogen exploiting the corresponding measurements in electronic hydrogen. The proton structure uncertainty in the calculation of $\alpha^5$ contribution is sizably reduced.
\end{abstract}

\maketitle

The theoretical knowledge of the two-photon exchange (TPE) correction to the hyperfine splitting (HFS) of the S energy levels in muonic hydrogen exceeds by two orders of magnitude the expected ppm level of the experimental accuracy in the forthcoming measurements of 1S HFS by CREMA~\cite{Pohl:2016tqq} and FAMU~\cite{Dupays:2003zz,Adamczak:2016pdb} collaborations as well as at J-PARC~\cite{Ma:2016etb}. In the ordinary hydrogen, the uncertainty of TPE is even six orders of magnitude above the experimental precision \cite{Carlson:2008ke,Eides:2000xc}, the measurements were performed in the 1970s  \cite{Hellwig:1970,Zitzewitz:1970,Essen:1971x,Morris:1971,Essen:1973,Reinhard:1974,Vanier:1976,Petit:1980,Cheng:1980} and discussed in Refs. \cite{Karshenboim:2005iy,Horbatsch:2016xx}.
\vspace{-0.2cm}
\begin{figure}[h]
\begin{center}%\centering
\includegraphics[width=0.225\textwidth]{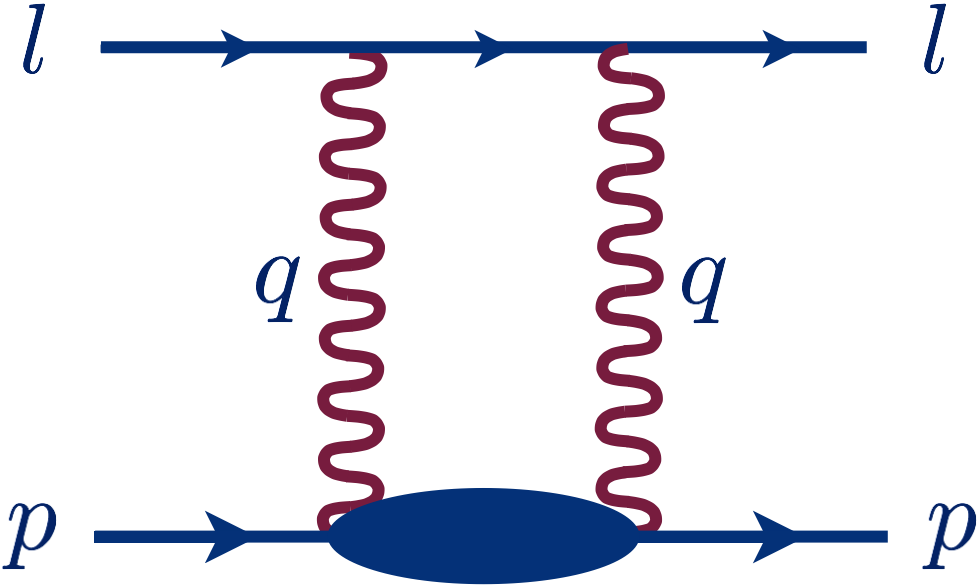}
\end{center}
\caption{Two-photon exchange graph.}
\label{TPE_graph}
\end{figure}

\vspace{-0.3cm}
The graph with two exchanged photons, see Fig. \ref{TPE_graph} for the notation of particles momenta, also contributes the largest theoretical uncertainty in the proton size extractions from the Lamb shift in muonic hydrogen  ($\mu \mathrm{H}$) \cite{Pohl:2010zza,Antognini:1900ns}. It was a subject of extensive theoretical studies \cite{Pachucki:1996zza,Faustov:1999ga,Pineda:2002as,Pineda:2004mx,Nevado:2007dd,Carlson:2011zd,Hill:2012rh,Birse:2012eb,Miller:2012ne,Alarcon:2013cba,Gorchtein:2013yga,Peset:2014jxa,Tomalak:2015hva,Caprini:2016wvy,Hill:2016bjv} since the formulation of the {\it proton radius puzzle} in 2010, when the accurate extraction of the proton charge radius ($\mathrm{R}_\mathrm{E}$) from the muonic hydrogen Lamb shift by the CREMA Collaboration at PSI \cite{Pohl:2010zza,Antognini:1900ns} gave a significantly smaller result than the electron data based extractions \cite{Bernauer:2010wm,Bernauer:2013tpr,Mohr:2012tt}, see Refs. \cite{Antognini:1900ns,Carlson:2015jba} for recent reviews.

Besides the Lamb shift, the CREMA Collaboration has extracted the HFS of the 2S energy level in $\mu \mathrm{H}$ \cite{Antognini:2013jkc}, where the leading theoretical uncertainty is also coming from TPE. The corresponding correction to HFS of S energy levels is expressed in terms of the proton elastic form factors and spin structure functions \cite{Zemach:1956zz,Iddings:1959zz,Iddings:1965zz,Drell:1966kk,Faustov:1966,Faustov:1970,Bodwin:1987mj,Faustov:2001pn,Martynenko:2004bt,Carlson:2008ke,Carlson:2011af,Hagelstein:2015egb,Peset:2016wjq,Tomalak:2017owk}. The first full dispersive calculation of this contribution was performed in Refs. \cite{Carlson:2008ke,Carlson:2011af}, where it was evaluated with 213 ppm uncertainty. The subsequent studies expressing the region with small photons virtualities in terms of proton radii and moments of the spin structure functions led to the uncertainty 105 ppm \cite{Tomalak:2017npu}. The model-independent evaluation within the frameworks of Non-Relativistic Quantum Electrodynamics  and Chiral Perturbation Theory exploiting the electronic hydrogen ($e\mathrm{H}$) HFS measurement was recently performed in Ref. \cite{Peset:2016wjq}, for results in Chiral Effective Field Theory see Ref.~\cite {Hagelstein:2015egb}. 

In this work, we aim to reduce the proton structure uncertainty in the dispersive evaluation of the TPE correction to HFS of the S energy levels in $\mu \mathrm{H}$ exploiting precise measurements of the HFS in $e\mathrm{H}$ \cite{Karshenboim:2005iy}.

The TPE contribution to the $n$S-level HFS $ \delta\mathrm{E}^{\mathrm{HFS}}_{n\mathrm{S}} $ is expressed in terms of the relative correction $ \Delta_{\mathrm{HFS}} $ and the leading-order HFS $ \mathrm{E}^{\mathrm{HFS},0}_{n\mathrm{S}} $ as \cite{Eides:2000xc} 
\ber
 \delta \mathrm{E}^{\mathrm{HFS}}_{n\mathrm{S}} & = & \Delta_{\mathrm{HFS}} \left(m \right) \mathrm{E}^{\mathrm{HFS},0}_{n\mathrm{S}} , \\
 \mathrm{E}^{\mathrm{HFS},0}_{n\mathrm{S}} & = & \frac{8}{3} \frac{ m_\mathrm{r} (m)^3 \alpha^4}{ M m} \frac{\mathrm{\mu}_\mathrm{P}}{n^3}, \label{LO_HFS}
\eer
where $M$ and $m$  are the proton and the lepton masses in  the energy units, $ m_\mathrm{r} (m) = M m / ( M + m )$ is the reduced mass, $ \mathrm{\mu}_\mathrm{P}  $ is the proton magnetic moment, $ \alpha $ is the electromagnetic coupling constant and $n = 1, 2, 3, ...$ is the principal quantum number. $ \Delta_{\mathrm{HFS}} $ is usually defined as a sum of the Zemach correction $ \Delta^{\mathrm{Z}} $, the recoil correction $ \Delta^{\mathrm{R}} $ and the polarizability correction $ \Delta^{\mathrm{pol}} $ \cite{Eides:2000xc,Tomalak:2017owk}:
\ber \label{traditional}
\Delta_{\mathrm{HFS}} & = & \Delta^{\mathrm{Z}} + \Delta^{\mathrm{R}} +  \Delta^{\mathrm{pol}},
\eer
which can be expressed as integrals over the photon energy $\nu_\gamma = \left( p \cdot q \right)/ M$ and the virtuality $Q^2 = -q^2$:
\ber \label{zemach_correction}
\Delta^{\mathrm{Z}}& = &\frac{8 \alpha m_\mathrm{r}}{\pi }  \int \limits^{\infty}_{0} \frac{\mathrm{d} Q}{Q^2} \left( \frac{\mathrm{G}_\mathrm{E}\left(Q^2\right) \mathrm{G}_\mathrm{M}\left(Q^2\right)  }{\mathrm{\mu}_\mathrm{P}}  - 1 \right), \\
\Delta^{\mathrm{R}}  & = &  \frac{\alpha}{\pi} \int \limits^{\infty}_{0} \frac{\mathrm{d} Q^2}{Q^2}  \frac{ \left( 2 + \rho\left(\tau_\mathrm{l}\right) \rho\left(\tau_\mathrm{P} \right)  \right) \mathrm{F}_\mathrm{D}\left(Q^2\right)  }{ \sqrt{\tau_\mathrm{P}} \sqrt{1+\tau_\mathrm{l}} + \sqrt{\tau_\mathrm{l}} \sqrt{1+\tau_\mathrm{P}} } \frac{ \mathrm{G}_\mathrm{M}\left(Q^2\right)  }{\mathrm{\mu}_\mathrm{P}} \nonumber \\ 
&+&  \frac{3 \alpha}{\pi} \int \limits^{\infty}_{0} \frac{\mathrm{d} Q^2}{Q^2}  \frac{ \rho\left(\tau_\mathrm{l}\right) \rho\left(\tau_\mathrm{P} \right) \mathrm{F}_\mathrm{P}\left(Q^2\right)  }{ \sqrt{\tau_\mathrm{P}} \sqrt{1+\tau_\mathrm{l}} + \sqrt{\tau_\mathrm{l}} \sqrt{1+\tau_\mathrm{P}} } \frac{ \mathrm{G}_\mathrm{M}\left(Q^2\right)  }{\mathrm{\mu}_\mathrm{P}}  \nonumber \\ 
&-& \frac{\alpha}{\pi} \int \limits^{\infty}_{0} \frac{\mathrm{d} Q}{Q}\left(\frac{m}{M} \frac{\beta_1 \left(\tau_\mathrm{l}\right) \mathrm{F}^2_\mathrm{P} \left(Q^2\right)}{\mathrm{\mu}_\mathrm{P}} - \frac{8 m_\mathrm{r}}{Q} \right) - \Delta_Z,  \label{recoil_correction} \\
\Delta^{\mathrm{pol}}  & = &  \frac{2 \alpha}{\pi \mathrm{\mu}_\mathrm{P}} \int \limits^{\infty}_{0} \frac{\mathrm{d} Q^2}{Q^2} \int \limits^{\infty}_{\nu^{\mathrm{inel}}_{\mathrm{thr}}} \frac{\mathrm{d} \nu_\gamma}{\nu_\gamma}  \frac{\left( 2 + \rho\left(\tau_\mathrm{l}\right) \rho\left(\tilde{\tau}\right) \right)  g_1 \left(\nu_\gamma, Q^2 \right)}{  \sqrt{\tilde{\tau}} \sqrt{1+\tau_\mathrm{l}} + \sqrt{\tau_\mathrm{l}} \sqrt{1+\tilde{\tau}}   } \nonumber \\
 &-& \frac{6 \alpha}{\pi \mathrm{\mu}_\mathrm{P}} \int \limits^{\infty}_{0} \frac{\mathrm{d} Q^2}{Q^2} \int \limits^{\infty}_{\nu^{\mathrm{inel}}_{\mathrm{thr}}} \frac{\mathrm{d} \nu_\gamma}{ \nu_\gamma} \frac{1}{\tilde{\tau}}  \frac{ \rho\left(\tau_\mathrm{l}\right) \rho\left(\tilde{\tau} \right)  g_2 \left(\nu_\gamma, Q^2 \right) }{  \sqrt{\tilde{\tau}} \sqrt{1+\tau_\mathrm{l}} + \sqrt{\tau_\mathrm{l}} \sqrt{1+\tilde{\tau}}   } \nonumber \\
&+& \frac{\alpha}{\pi }  \int \limits^{\infty}_{0} \frac{\mathrm{d} Q}{Q} \frac{m}{M}\frac{\beta_1 \left(\tau_\mathrm{l}\right) \mathrm{F}^2_\mathrm{P} \left(Q^2\right)}{\mathrm{\mu}_\mathrm{P}} , \label{polar_correction}
\eer
with $ \beta_1(\tau) = \rho(\tau)^2 - 4 \rho(\tau)$, $\rho(\tau) = \tau - \sqrt{\tau ( 1 + \tau )}$ and
\ber
 \tau_\mathrm{l} = \frac{Q^2}{4 m^2}, ~~~~~ \tau_\mathrm{P} = \frac{Q^2}{4 M^2}, ~~~~~  \tilde{\tau} = \frac{\nu_\gamma^2}{Q^2}.  \label{taus}
\eer
 The proton structure enters the TPE correction through the Dirac, Pauli, Sachs electric and magnetic form factors $ \mathrm{F}_\mathrm{D} (Q^2) $, $ \mathrm{F}_\mathrm{P}(Q^2) $, $ \mathrm{G}_\mathrm{E} (Q^2) $ and $ \mathrm{G}_\mathrm{M}(Q^2) $ respectively as well as through the spin-dependent inelastic proton structure functions $g_1 \left(\nu_\gamma, Q^2 \right)$ and $g_2 \left(\nu_\gamma, Q^2 \right)$. The photon energy integration starts from the pion-nucleon inelastic threshold $ \nu^{\mathrm{inel}}_{\mathrm{thr}} = m_\pi +  \left( m_\pi^2 + Q^2 \right) / \left( 2 M \right)$, where $ m_{\pi} $ denotes the pion mass.

We propose to improve the theoretical prediction of the TPE correction in $\mu \mathrm{H}$ $ \Delta^{\mathrm{impr}}_{\mathrm{HFS}} \left( m_\mu \right) $ from the known correction in $e \mathrm{H}$ $ \Delta^{\mathrm{exp}}_{\mathrm{HFS}} \left( m_e \right)$:
\ber \label{XXX}
\Delta^{\mathrm{impr}}_{\mathrm{HFS}} \left(m_\mu \right) & = &\frac{m_\mathrm{r}( m_\mu)}{m_\mathrm{r}( m_e)} \Delta^{\mathrm{exp}}_{\mathrm{HFS}} \left( m_e \right) \nonumber \\ 
&+& \Delta_{\mathrm{HFS}} \left( m_\mu \right) - \frac{m_\mathrm{r}( m_\mu)}{m_\mathrm{r}( m_e)} \Delta_{\mathrm{HFS}} \left( m_e \right),
\label{TPE_expression}
\eer
performing the photon virtuality integration for the ansatz in the last line of Eq. (\ref{TPE_expression}) as a whole. In contrast to Ref. \cite{Peset:2016wjq}, we do not expand the correction to the hyperfine splitting in lepton mass and evaluate the difference in Eq. (\ref{TPE_expression}), which is weighted by the reduced mass but not by the lepton mass itself. Introducing the ratio $m_\mathrm{r}(m_\mu)/m_\mathrm{r}(m_e)$ in Eq. (\ref{XXX}), we exactly \footnote{The Zemach correction also cancels in Ref. \cite{Peset:2016wjq}, where it is defined for the infinitely heavy proton, i.e. $m_r$ is replaced by $m$ in Eq. (\ref{zemach_correction}). Accounting for the lepton mass in Eq. (\ref{zemach_correction}), it cancels in Ref. \cite{Peset:2016wjq} at the leading order in the lepton mass expansion.} cancel the Zemach contribution of Eq. (\ref{zemach_correction}), which is a main source of the theoretical uncertainty due to the pure knowledge of the proton electromagnetic form factors and radii \cite{Tomalak:2017npu}. As we will see in the following, the main source of the uncertainty coming from the errors of the proton spin structure functions, the polarizability correction of Eq. (\ref{polar_correction}), also scales as a reduced mass within errors producing a small number in the weighted difference of Eq. (\ref{XXX}). As a result, the corresponding uncertainty is smaller than the uncertainty of $ \Delta^{\mathrm{pol}} $ from the direct evaluation of integrals.

We determine the polarizability correction by the method of Ref. \cite{Tomalak:2017npu} expressing it in terms of the first moment $\mathrm{I}_1 \left( Q^2 \right)$ of the proton spin structure function $ g_1 \left(\nu_\gamma, ~Q^2\right)$:
\ber
\mathrm{I}_1\left( Q^2 \right) & =& \int \limits^{\infty}_{ \nu^{\mathrm{inel}}_{\mathrm{thr}}} g_1 \left(\nu_\gamma, ~Q^2\right) \frac{M \mathrm{d} \nu_\gamma}{\nu_\gamma^2},
\eer
which at $Q^2 = 0 $ reduces to the Gerasimov-Drell-Hearn sum rule \cite{Drell:1966jv,Gerasimov:1965et}:
\ber
\mathrm{I}_1(0) & = & - \frac{\left(\mathrm{\mu}_\mathrm{P} - 1 \right)^2}{4}. \label{I1_moment}
\eer
For the polarizability contribution, we expand $\mathrm{I}_1 \left( Q^2 \right)$ with the low-energy constant $ {\mathrm{I}}_1\left( 0 \right)' = 7.6 \pm 2.5 ~\mathrm{GeV}^{-2} $ \cite{Prok:2008ev} up to $ Q^2_0 = 0.0625~\mathrm{GeV}^2$. We additionally account for the errors due to the choice of the splitting parameter $Q_0$ and due to the contribution of higher terms in $Q^2$ expansion \cite{Tomalak:2017npu}.

For the recoil TPE, we exploit the parametrization of the elastic form factors from Refs. \cite{Bernauer:2010wm,Bernauer:2013tpr}, which is based on the unpolarized and polarization transfer world data. The proton spin structure functions parametrization is based on Refs. \cite{Prok:2008ev,Kuhn:2008sy,griffioen,Sato:2016tuz,Fersch:2017qrq}. We calculate the error adding uncertainties from the form factors and spin structure functions under the Q-integration in Eq. (\ref{XXX}) in quadrature.

\begin{figure}[h]
\begin{center}%\centering
\includegraphics[width=0.33\textwidth]{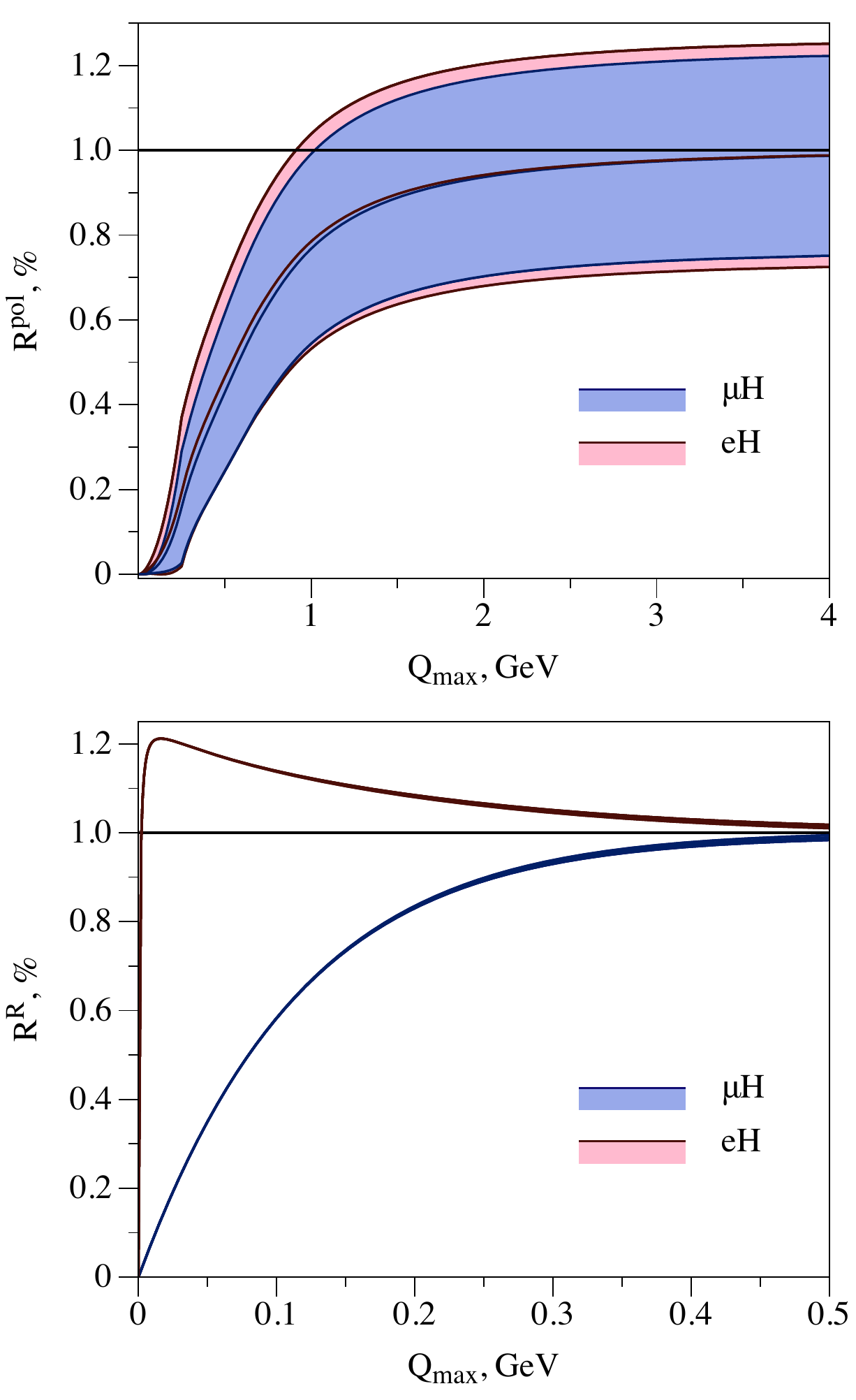}
\end{center}
\caption{The saturation of the polarizability and recoil contributions to the $n$S-level HFS in $ \mu \mathrm{H} $  and $e \mathrm{H}$.}
\label{integrand}
\end{figure}

In Fig. \ref{integrand}, we study the saturation of the different contributions $ \Delta^\mathrm{i}$ to the HFS correction of Eqs. (\ref{recoil_correction}, \ref{polar_correction}) in $e \mathrm{H}$ and $\mu \mathrm{H}$:
\ber
 \Delta^\mathrm{i} \left( Q_\mathrm{max} \right)=  \int \limits_0^{ Q_\mathrm{max}} \mathrm{I}^{\mathrm{i}}(Q) \mathrm{d} Q, 
\eer
where we present the following ratio for the recoil and polarizability contributions $ R^\mathrm{i} \left( Q_\mathrm{max} \right)=  \Delta^\mathrm{i} \left( Q_\mathrm{max} \right) /\Delta^\mathrm{i} $ as a function of the integral cutoff $ Q_\mathrm{max}$. \footnote{Note that $ \mathrm{I}^{\mathrm{pol}}(0) = 0$,  $ \mathrm{I}^{\mathrm{R}}(0) = \frac{\alpha}{\pi} m_r \frac{ 2 ( M + m ) +  \mathrm{\mu}_\mathrm{P} M }{m} \frac{\mathrm{\mu}_\mathrm{P} - 1}{\mathrm{\mu}_\mathrm{P} } \frac{1}{M^2}$ and $ \mathrm{I}^{\mathrm{Z}}(0) = - \frac{4}{3} \frac{\alpha}{\pi} m_r \left( \mathrm{R}^2_\mathrm{E} + \mathrm{R}^2_\mathrm{M} \right) $, with the magnetic radius $\mathrm{R}_\mathrm{M}$.}

The behavior of the polarizability correction in $ \mu \mathrm{H} $  and $e \mathrm{H}$ is very similar. As a result, the corresponding contribution to $\Delta^{\mathrm{impr}}_{\mathrm{HFS}} \left(m_\mu \right) $ is close to zero with the uncertainty exceeding the central value: \footnote{Varying the lepton mass between the electron and muon values, the relative change of $ \Delta^\mathrm{pol}(m)/m_r(m)$ is less than $3~\%$.}
\beq \label{polar}
 \Delta^{\mathrm{pol}} \left( m_\mu \right) -\frac{m_\mathrm{r}( m_\mu)}{m_\mathrm{r}( m_e)} \Delta^{\mathrm{pol}} \left( m_e \right) =  9.4 \pm 19.1~\mathrm{ppm}.
\eeq

The saturation of the recoil correction is qualitatively different. In $\mu \mathrm{H}$, the integrand $\mathrm{I}^\mathrm{R}(Q)$ has a definite positive sign. While in 
$e \mathrm{H}$ the integrand changes sign at $ Q\sim 0.016~\mathrm{GeV}$, which is driven by the kinematical prefactor and is sensitive mainly to the proton magnetic moment. The recoil correction $ \Delta^\mathrm{R}$ has a relatively small error around $ 6-8~\mathrm{ppm}$ in $ \mu \mathrm{H}$~\cite{Tomalak:2017owk,Tomalak:2017npu}. Consequently, the uncertainty of this contribution in Eq. (\ref{TPE_expression}):
\ber
 \Delta^{\mathrm{R}} \left( m_\mu \right) -\frac{m_\mathrm{r}( m_\mu)}{m_\mathrm{r}( m_e)} \Delta^{\mathrm{R}} \left( m_e \right) =  -143.1 \pm 3.2~\mathrm{ppm},
\eer
has the same order of magnitude.

Adding the recoil and polarizability contributions, we obtain the resulting proton structure correction:
\beq \label{total}
 \Delta_{\mathrm{HFS}} \left( m_\mu \right) - \frac{m_\mathrm{r}( m_\mu)}{m_\mathrm{r}( m_e)}  \Delta_{\mathrm{HFS}} \left( m_e \right) = -133.6 \pm 20.0~\mathrm{ppm}.
\eeq

% Polarizability correction 0.8 pmm comes from large Q^2 > 10 GeV^2.
We extract the relative difference between the experimental value $ \mathrm{E}^{\mathrm{HFS},\mathrm{exp}}_{\mathrm{1S}} $ of the 1S HFS in $e\mathrm{H}$ \cite{Hellwig:1970,Zitzewitz:1970,Essen:1971x,Morris:1971,Essen:1973,Reinhard:1974,Vanier:1976,Petit:1980,Cheng:1980,Karshenboim:2005iy,Horbatsch:2016xx} and the theory prediction $\mathrm{E}^{\mathrm{HFS},\mathrm{QED}}_{\mathrm{1S}}$, which was shifted by well-known nonrecoil QED contributions from Ref. \cite{Eides:2000xc}:
\ber
\frac{\mathrm{E}^{\mathrm{HFS},\mathrm{exp}}_{\mathrm{1S}} - \mathrm{E}^{\mathrm{HFS},\mathrm{QED}}_{\mathrm{1S}}}{\mathrm{E}^{\mathrm{HFS},0}_{\mathrm{1S}}} =  -32.6170 \pm 0.0032~\mathrm{ppm}.
 \eer
 The extraction error is dominated by the uncertainty of the proton magnetic moment in the leading-order HFS $\mathrm{E}^{\mathrm{HFS},0}_{\mathrm{1S}}$ of Eq. (\ref{LO_HFS}). Furthermore, we account for the recoil and nuclear size corrections beyond the leading $ \alpha $ order as well as for the weak interaction contribution \cite{Karshenboim:1996ew,Eides:2000xc} that provide in total $ 0.085 \pm 0.027~\mathrm{ppm}$. We account also for the contribution of the axial-vector mesons $ -0.073 \pm 0.019~\mathrm{ppm}$ following the evaluation of this correction in $\mu \mathrm{H}$ \cite{Dorokhov:2017nzk}. Assuming the absence of other important contributions to the hydrogen hyperfine structure, we estimate the TPE correction $ \Delta^{\mathrm{exp}}_{\mathrm{HFS}} \left( m_e \right) $  to HFS as a remaining difference between theory and experiment: $ \Delta^{\mathrm{exp}}_{\mathrm{HFS}} \left( m_e \right) =  -32.629 \pm 0.033~\mathrm{ppm}$, and predict the corresponding TPE effect in $\mu \mathrm{H}$:
 \ber \label{correction1}
 \Delta^{\mathrm{impr}}_{\mathrm{HFS}} \left(m_\mu \right) =  -6201 \pm 20~\mathrm{ppm},
 \eer
 where the uncertainties in Eq. (\ref{XXX}) are added in quadrature. Assuming the relative contribution of higher orders to be suppressed by a factor of $\alpha$, which increases the uncertainty of $\Delta^{\mathrm{exp}}_{\mathrm{HFS}} \left( m_e \right)$ by an order of magnitude, i.e. $  \Delta^{\mathrm{exp}}_{\mathrm{HFS}} \left( m_e \right) =  -32.629 \pm 0.240~\mathrm{ppm}$, we obtain:
  \ber \label{correction2}
 \Delta^{\mathrm{impr}}_{\mathrm{HFS}} \left(m_\mu \right) = - 6201 \pm 49~\mathrm{ppm}.
 \eer
  
\begin{figure}[H]
\begin{center}%\centering
\includegraphics[width=0.42\textwidth]{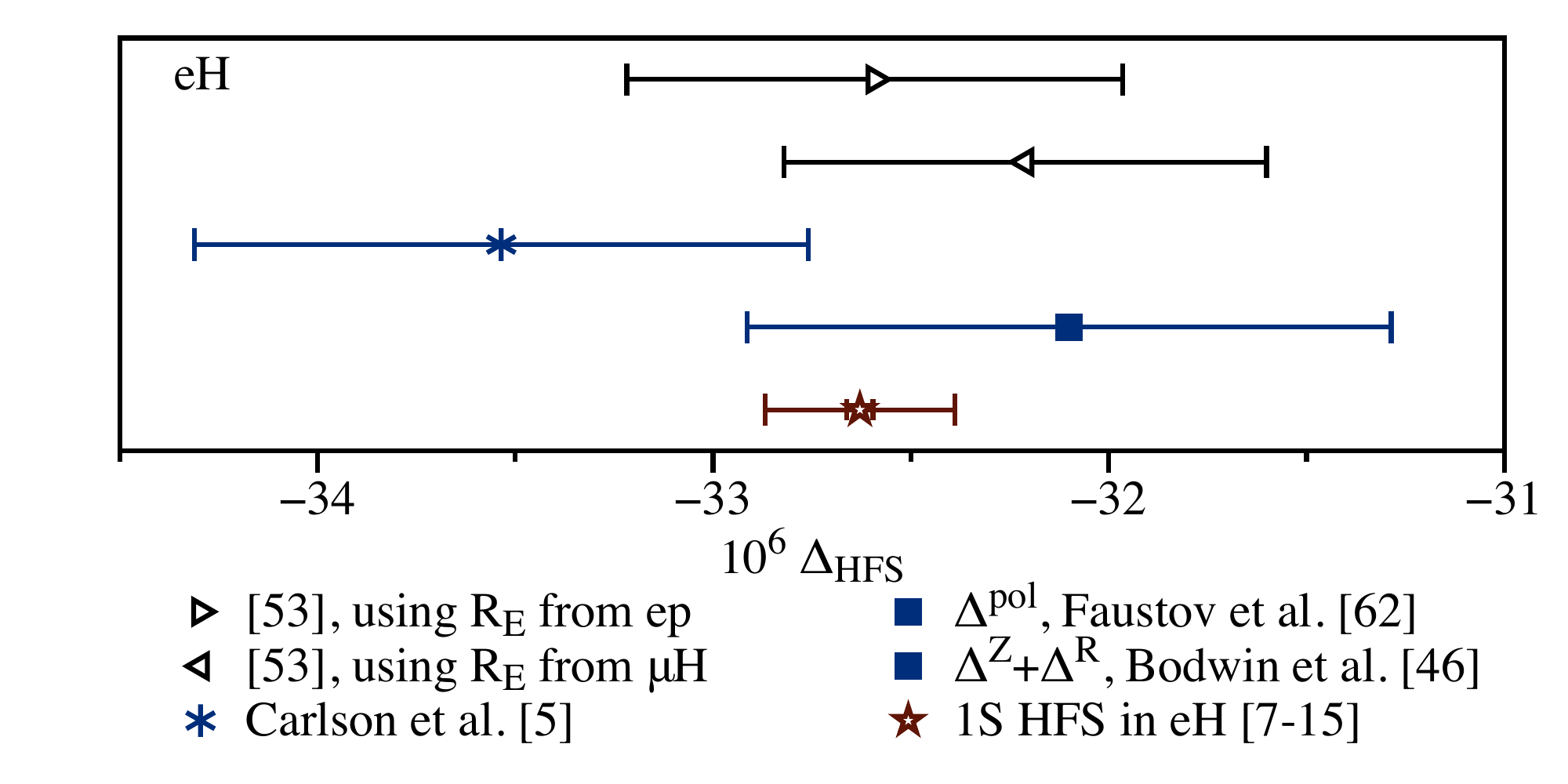}
\end{center}
\caption{Phenomenological extraction of the TPE correction to the $n$S-level HFS in $ e \mathrm{H} $ is compared with theoretical estimates. The Zemach and recoil corrections of Ref. \cite{Bodwin:1987mj} are combined with the polarizability contribution of Ref. \cite{Faustov:2002yp}. Results are presented in the chronological order starting from the lowest estimate.}
\label{comparisoneH}
\begin{center}%\centering
\includegraphics[width=0.42\textwidth]{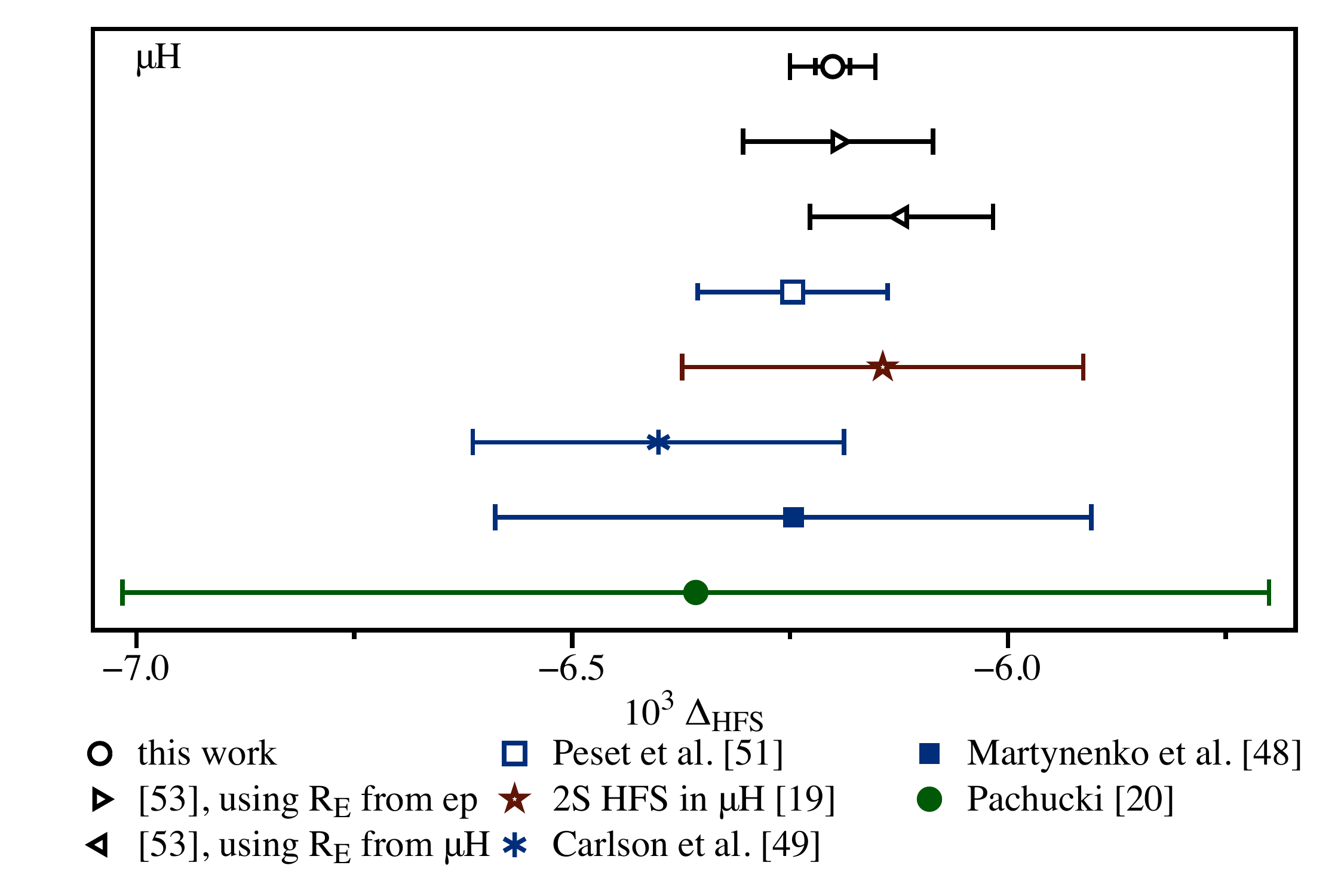}
\end{center}
\caption{TPE correction to the $n$S-level HFS in $ \mu \mathrm{H} $ of this work in comparison with other theoretical estimates and phenomenological extraction from the 2S HFS in $\mu\mathrm{H}$. Results are presented in the chronological order starting from the lowest estimate.}
\label{comparison}
\end{figure}

As a cross-check of the TPE estimate from the electronic hydrogen, we present a good agreement between our phenomenological extraction and the calculation for the ordinary hydrogen by the method of Ref. \cite{Tomalak:2017npu} (exploiting the magnetic radius value from Ref. \cite{Bernauer:2013tpr}) in Fig. \ref{comparisoneH}. We compare our evaluation with previous theoretical results and phenomenological extraction from the 2S HFS measurement \cite{Antognini:1900ns} in Fig. \ref{comparison}, where we also account for the radiative corrections of Refs. \cite{Faustov:2017hfo,Dorokhov:2017gst,Dorokhov:2017nzk}.
Our result is in a reasonable agreement with estimates of Refs. \cite{Pachucki:1996zza,Faustov:2001pn,Martynenko:2004bt,Carlson:2011af,Peset:2016wjq,Tomalak:2017owk,Tomalak:2017npu}, where we subtract the recoil effect of order $ \alpha^2$, the radiative correction to the Zemach contribution \cite{Carlson:2008ke} and account for the convention conversion correction of Ref. \cite{Carlson:2011af}. Our result and the TPE extraction from the 2S HFS measurement in $\mu \mathrm{H}$ \cite{Antognini:1900ns} are consistent within the error bands.
 
We convert the radiative corrections of Refs. \cite{Ivanov:1996ew,Martynenko:2004bt} to 1S energy level with account of recent evaluations in Refs. \cite{Peset:2016wjq,Dorokhov:2017gst,Faustov:2017hfo,Dorokhov:2017nzk}. We also calculate the hadronic vacuum polarization contirbutions exploiting up-to-date fits of the electron-proton annihilation cross section to hadrons \cite{Jegerlehner:2011mw,Jegerlehner:2017pr}. Assuming that there are no other contributions which are marginal in $e \mathrm{H}$ and can be amplified in $\mu \mathrm{H}$, we obtain the absolute value of the hyperfine splitting energy $ \mathrm{E}^{\mathrm{HFS}, \mu \mathrm{H}}_{\mathrm{1S}}$ and the corresponding frequency $\mathrm{\nu}^{\mathrm{HFS}, \mu \mathrm{H}}_{\mathrm{1S}}$:
 \ber
 \mathrm{E}^{\mathrm{HFS}, \mu \mathrm{H}}_{\mathrm{1S}} &=& 182.601 \pm 0.013~\mathrm{meV}, \\
 \mathrm{\nu}^{\mathrm{HFS}, \mu \mathrm{H}}_{\mathrm{1S}} &=& 44152.8 \pm 3.2~\mathrm{GHz}.
 \eer
We added the additional error $ \alpha  \delta \mathrm{E}^{\mathrm{HFS}}_{1\mathrm{S}}$ due to the possible contribution of higher orders. The hyperfine splitting of the 2S energy level in $\mu \mathrm{H}$ is given by
 \ber
 \mathrm{E}^{\mathrm{HFS}, \mu \mathrm{H}}_{\mathrm{2S}} &=& 22.8102 \pm 0.0016~\mathrm{meV}, \\
 \mathrm{\nu}^{\mathrm{HFS}, \mu \mathrm{H}}_{\mathrm{2S}} &=& 5515.49 \pm 0.40~\mathrm{GHz}.
 \eer
 
 ~
 
The knowledge of the HFS in $e \mathrm{H}$ allowed to pin down the proton structure uncertainty of the TPE contribution to $n$S energy levels in $\mu \mathrm{H}$. The error is given mainly by a poor knowledge of the low-energy constant $\mathrm{I}_1(0)'$ as well as uncertainties in the proton spin structure functions $ g_1 $ and $g_2 $. The proton spin structure studies at JLab \cite{Zheng:2009zza,SANE:2011aa,Zielinski:2017gwp} will allow to reduce the uncertainty further. The relation between HFS in  $e \mathrm{H}$ and $\mu \mathrm{H}$ provides an empirical test of the applied radiative corrections. The obtained result could help to adjust the laser frequency in measurements of the 1S HFS in $\mu \mathrm{H}$ with $ \mathrm{ppm} $ precision level \cite{Pohl:2016tqq,Dupays:2003zz,Ma:2016etb,Adamczak:2016pdb}.

\newpage

We thank Randolf Pohl and Marc Vanderhaeghen for reading this manuscript, valuable discussions about the HFS measurements and useful comments. We acknowledge the communication with Alexei Martynenko regarding the recent updates in radiative corrections. This work was supported by the Deutsche Forschungsgemeinschaft (DFG) through Collaborative Research Center ``The Low-Energy Frontier of the Standard Model'' (SFB 1044).

\end{document}